\pdfoutput=1
\documentclass[manuscript, screen]{acmart}
\AtBeginDocument{%
  \providecommand\BibTeX{{%
    \normalfont B\kern-0.5em{\scshape i\kern-0.25em b}\kern-0.8em\TeX}}}

\setcopyright{acmlicensed}
\copyrightyear{2024}
\acmYear{2024}
\acmDOI{XXXXXXX.XXXXXXX}

\usepackage{algorithm}
\usepackage{algpseudocodex}

\usepackage{amsmath,amsfonts}
\usepackage{amstext}
\usepackage{stmaryrd}

\usepackage{float}

\usepackage[title]{appendix}

\usepackage{tikz,tikzscale}
\usepackage{graphicx}

\usepackage{pgfplots}
\usepgfplotslibrary{groupplots}

\usepackage{booktabs}

\usepackage{listings}

\usepackage{mathtools}

\usepackage[mode=multiuser,status=draft]{fixme} 
\fxsetup{innerlayout=inline}
\fxsetup{targetlayout=colorcb}
\fxusetheme{color}
\FXRegisterAuthor{cc}{envcc}{Author}
\FXRegisterAuthor{gk}{envgk}{GK}
\FXRegisterAuthor{rw}{envre}{\textbf{R}}

\begin{document}

\title{Acceleration of Tensor-Product Operations with Tensor Cores}

\author{Cu Cui}
\authornote{Submitted to the editors 12.04.2024.}
\email{cu.cui@iwr.uni-heidelberg.de}
\affiliation{%
  \department[1]{Interdisciplinary Center for Scientific Computing (IWR)}
  \institution{Heidelberg University}
  \streetaddress{Im Neuenheimer Feld 205}
  \postcode{69120}
  \city{Heidelberg}
  \country{Germany}
}


\begin{abstract}
    In this paper, we explore the acceleration of tensor product operations in finite element methods, leveraging the computational power of the NVIDIA A100 GPU Tensor Cores. We provide an accessible overview of the necessary mathematical background and discuss our implementation strategies. Our study focuses on two common programming approaches for NVIDIA Tensor Cores: the C++ Warp Matrix Functions in \texttt{nvcuda::wmma} and the inline Parallel Thread Execution (PTX) instructions \texttt{mma.sync.aligned}. A significant focus is placed on the adoption of the versatile inline PTX instructions combined with a conflict-free shared memory access pattern, a key to unlocking superior performance. When benchmarked against traditional CUDA Cores, our approach yields a remarkable 2.3-fold increase in double precision performance, achieving 8 TFLOPS/s—45\% of the theoretical maximum. Furthermore, in half-precision computations, numerical experiments demonstrate a fourfold enhancement in solving the Poisson equation using the flexible GMRES (FGMRES) method, preconditioned by a multigrid method in 3D. This is achieved while maintaining the same discretization error as observed in double precision computations. These results highlight the considerable benefits of using Tensor Cores for finite element operators with tensor products, achieving an optimal balance between computational speed and precision.
\end{abstract}




\maketitle

\section{Introduction}

In high-performance computing, many-core processors have become ubiquitous, particularly in powering numerous TOP500 supercomputers. Eight out of the top ten supercomputers (according to the HPL benchmark) employ many-core accelerators, seven of which are GPUs~\cite{top500June23}. This surge in GPU adoption underscores a significant shift in computational strategies, especially in the domain of high-order finite element methods (FEM) with their high computational intensity. These methods are now routinely deployed on parallel architectures scaling up to millions of CPU cores. Hence, the development of ﬁnite element codes achieving a high level of utilization of these devices is still an active ﬁeld. In this work we demonstrate the efficient use of the NVIDIA Tensor Cores to accelerate the evaluation of finite element operations.

The optimization of FEM operations has evolved significantly. Early studies dating back to 2005 focused on employing GPU accelerators for solving elliptic Partial Differential Equations (PDEs) in two-dimensional domains~\cite{goddeke2005accelerating,goddeke2007exploring}. Subsequent research expanded into more complex applications, including Navier–Stokes solvers and higher order methods~\cite{goddeke2009gpu,klockner2009nodal}. A critical observation in these developments was the inefficiency of matrix-based schemes, particularly at higher polynomial degrees due to dense coupling between degrees of freedom (DoF). This inefficiency led to a paradigm shift towards matrix-free approaches, where the matrix-vector product in iterative solvers is replaced by on-the-fly evaluation of discretized differential operators~\cite{KronbichlerKormann12,KronbichlerKormann19}. Many high-performance implementations with tensorial evaluation by specific tuning techniques were developed for GPUs using traditional CUDA cores~\cite{KronbichlerLjunqkvist19,ljungkvist2017matrix,remacle2016gpu,Cui2023}. For instance, the NekRS, a GPU-based incompressible fluid flow simulator, has achieve unprecedented simulation scales and accuracy on high-performance computing platforms~\cite{Abdelfattah2021,Kolev2021,Merzari2023}. Studies~\cite{swirydowicz2019acceleration,KARAKUS2019380} have demonstrated near-roofline performance for CUDA Core kernel optimization in FEM local operators. The trend towards higher on-chip parallelism emphasizes the importance of fine-grained parallel execution of higher-order operators on newer-generation units like Tensor Cores.

NVIDIA's Tensor Cores, introduced with the Volta Architecture~\cite{V100}, marked a significant milestone in GPU computing. Originally developed for deep learning applications, these cores leverage their low- and mixed-precision capabilities for a range of computational tasks~\cite{dakkak2019accelerating,finkelstein2022quantum,haidar2018harnessing,li2021tcfft}. NVIDIA's optimized libraries like cuBLAS~\cite{cublas} and CUTLASS~\cite{cutlass} have further facilitated the adoption of Tensor Cores. 
Despite previous work demonstrating significant benefits of using Tensor Cores for various non-matrix multiplicative tasks~\cite{Wan2022,Benoit2022,Ji2021,Ji2022}, no current studies have explored their application to accelerating tensor product operations in finite element computations.

Recent work by \cite{ruda2022} discusses a finite element Poisson solver on lower precision accelerator hardware, but is restricted to matrix-based methods.
In this paper, we delve into the implementation of finite element operators using a matrix-free approach, harnessing the newly introduced flexible \texttt{mma} instructions of Tensor Cores. Our research compares the performance of Tensor Cores with traditional CUDA Cores and explores specific optimization techniques related to shared memory. By combining the flexibility of inline PTX instructions with carefully designed shared memory access patterns, our conflict-free approach improves performance by more than a factor of two.
To evaluate the performance of the computational kernel, we use an empirical roofline model along with a model that takes shared memory into account~\cite{swirydowicz2019acceleration}. With our optimized kernel, the performance is close to the roofline performance.
Additionally, we investigate mixed-precision methods~\cite{goddeke2007performance, KronbichlerLjunqkvist19}, employing half-precision multigrid V-cycle as a preconditioner and double-precision FGMRES for outer iterations. Our numerical experiments reveal a fourfold increase in solving linear systems' speed using these methods without compromising accuracy.

The remainder of this article is structured as follows. We begin with a mathematical description of the model problem, employing the discontinuous Galerkin (DG) method. This is followed by an introduction to the Tensor Core programming model used in our kernel designs. We then detail the implementation and optimization strategies through two benchmark problems. The efficiency of two cores in solving Poisson problems is examined. Finally, we conclude with remarks.

\section{Problem description}\label{sec:model}

In this study, we address the Poisson equation, a fundamental model in scientific computing, and demonstrate its solution using tensor cores. The Poisson equation is given by:
\begin{gather}\label{eq:poisson}
\begin{aligned}
    -\Delta u &= f & \text{ in } & \Omega \\
    u &{} = g & \text{ on } &\partial\Omega,
\end{aligned}
\end{gather}
$f$ and $g$ are given functions in $L^2(\Omega)$ and $L^2(\partial\Omega)$, respectively.
This equation, subject to Dirichlet boundary conditions, is defined over a domain $\Omega \subset \mathbb{R}^3$.
To discretize the problem, we employ the symmetric interior penalty method (SIPG), as detailed in~\cite{arnold1982interior,arnold2002unified}. To this end, we 
subdivide $\Omega$ into a mesh $\mathcal{T}_h$, comprising hexahedral cells $K$. These cells are mapped from a reference cell $\widehat{K} = [0, 1]^3$ using a transformation $F_K$. The discontinuous, tensor product polynomials $\mathbb{Q}_k$ form the basis of our shape function space $V(\widehat{K})$ on $\widehat{K}$. 
This space is constructed from tensor products of Lagrangian interpolation polynomials of degree $k$, associated with Gauss-Lobatto points. By composing these with the mapping $F_K$, we obtain the shape function spaces $V(K)$ for each grid cell, where $\phi_{K,i}(x) = \widehat{\phi}_i(F^{-1}_K(x))$. The finite element space is then defined as:
\begin{equation*}
    V_h \coloneqq \{v \in L^2(\Omega) | v_K \in V(K) \text{ for all } K \in \mathcal{T}_h\}.
\end{equation*}
For any interior interface between two cells $K^{+}$ and $K^{-}$ in the set $\mathcal{E}_{h}^{\circ}$, we define traces of functions $v \in V_{h}$ on $e \in \mathcal{E}_{h}^{\circ}$ from $K^{ \pm}$ as $v^{ \pm}$. Here, the averaging operator is introduced:
\begin{equation*}
\{v\}(\boldsymbol{x})=\frac{1}{2}\left(v^{+}(\boldsymbol{x})+v^{-}(\boldsymbol{x})\right), \quad \boldsymbol{x} \in e .
\end{equation*}
For boundary faces ($e \in \varepsilon_{h}^{\partial}$), the operator simplifies to $\{v\}(\boldsymbol{x})=v(\boldsymbol{x})$. Employing $\boldsymbol{n}$ as the outward normal of cell $K$ at face $e$, the discretization of our model problem via SIPG is: find $u_h \in V_h$, such that
\begin{equation}\label{eq:bilinear_form}
\begin{aligned}
& \int_{\mathcal{T}_{h}} \nabla u_h \cdot \nabla v \,d\boldsymbol{x} +  \int_{\mathcal{E}_{h}}\left(\gamma_e\{u_h \boldsymbol{n}\} \cdot\{v \boldsymbol{n}\}-\{\nabla u_h\} \cdot\{v \boldsymbol{n}\}-\{u_h \boldsymbol{n}\} \cdot\{\nabla v\}\right) d \sigma(\boldsymbol{x}) \\ 
& = \int_{\mathcal{T}_{h}} fv \,d \boldsymbol{x} \quad \text{on } v \in V_h,
\end{aligned}
\end{equation}
where $\gamma_e$ represents the edge-wise penalty parameter, calculated as $\gamma_e=k(k+1)\left(\frac{1}{h^{+}}+\frac{1}{h^{-}}\right)$.

\subsection{Matrix-free implementation of discontinuous Galerkin finite element operator}\label{sec:mf_operator}

The matrix-free evaluation of a finite element operator $v=A u=\left(\sum_{K \in \mathcal{T}_h} P_K^{\top} A_K P_K\right) u$ is done by a loop over all cells in the mesh as follows:
\begin{itemize}
    \item[(i)] Initialization: Set vector $v=0$
    \item[(ii)] Cell Loop: Iterate over each cell in the mesh
    \begin{itemize}
        \item[(a)] Gather local vector values: For each cell $K$, compute $u_K=P_K u$
        \item[(b)] Cell Operation: Apply operation $v_K=A_K^{c} u_K$ (without forming $A_K^c$)
        \item[(c)] Face Loop: For each of the $2d$ faces of the current cell, do
        \begin{itemize}
            \item[($\alpha$)] Face Operation: Update $v_K$ by applying the face operation, $v_K = v_K + A_K^f u_K$
        \end{itemize}
        \item[(d)] Global Summation: Sum the results into the global solution vector: $v=v+P_K^{\top} v_K$
\end{itemize}
\end{itemize}
In this framework, the matrix $P_K$ denotes the mapping of index vector  between global vector entries and the ones on the cell. $A_K^{c}$ and $A_K^{f}$ are the local cell and face integrals defined in~\eqref{eq:bilinear_form}, respectively. The evaluation of these integrals is performed by numerical quadrature without explicitly constructing the matrix $A_K$. For illustrative purposes, consider the matrix-free implementation of the local stiffness matrix $A_K^c$ on a hexahedral element $K$ in three dimensions:
\begin{equation}
    A_{i j}^K=\int_K \nabla \phi_i \cdot \nabla \phi_j d x,
\end{equation}
Transformed to the reference cube via mapping $F_K$, this becomes:
\begin{equation}
    A_{i j}^K=\int_{\widehat{K}} \nabla \phi_i^T\left(J^K\right)^{-T}\left|J^K\right|\left(J^K\right)^{-1} \nabla \phi_j d \boldsymbol{\xi}
\end{equation}
where $J^K$ denotes the Jacobian of the transformation from unit to real cell. 
Choosing the basis functions as tensor products of 1D Lagrange interpolating polynomials defined on $N$ Gauss-Lobatto points, we have
\begin{equation}
\phi_{i j k}(\xi)=\phi_i\left(\xi_1\right) \phi_j\left(\xi_2\right) \phi_k\left(\xi_3\right) \text {, }
\end{equation}
where $0 \leq i, j, k \leq p$. This choice enables the alignment of $n_q^d$ tensor product quadrature points with the node positions of the Lagrange polynomials, facilitating efficient evaluation of integrals. We express the quadrature weights and points using multi-index notation: $\left\{w_{i_q j_q k_q}\right\}_{i_q, j_q, k_q=1}^{n_q}$ and $\left\{\xi_{i_q j_q k_q}\right\}_{i_q, j_q, k_q=1}^{n_q}$. Consequently, the local stiffness matrix $A_{ij}^K$ is computed as~\cite{KronbichlerKormann12,KronbichlerKormann19}:
\begin{equation}
\begin{aligned}
A_{i j}^K= \sum_{i_q, j_q, k_q=1}^{n_q} w_{i_q j_q k_q} \nabla \phi_i^T\left(\xi_{i_q j_q k_q}\right)\left(J^K\left(\xi_{i_q j_q k_q}\right)\right)^{-T}  \left|J^K\left(\xi_{i_q, j_q k_q}\right)\right|\left(J^K\left(\xi_{i_q j_q k_q}\right)\right)^{-1} \nabla \phi_j\left(\xi_{i_q j_q k_q}\right) .
\end{aligned}
\end{equation}
As an aside, the solution $u$ at a quadrature point is then given by:
\begin{equation}
\begin{aligned}
 u\left(\xi_{i_q j_q k_q}\right)=\sum_{i, j, k=0}^p u_{i j k} \phi_{i j k}\left(\xi_{i_q j_q k_q}\right) =\sum_{k=0}^p \phi_k\left(\xi_{k_q}\right) \sum_{j=0}^p \phi_j\left(\xi_{j_q}\right) \sum_{i=0}^p u_{i j k} \phi_i\left(\xi_{i_q}\right)
\end{aligned}
\end{equation}
We denote by $S_i$ the $N \times N$ matrix of all $N$ one-dimensional shape functions $\phi^{1D}$ of degree $k = N - 1$, evaluated at $N$ quadrature points. $D_i$ represents the matrix of their derivatives along direction $i$.

With these formulations, we recast the local operator as a Kronecker product of local 1D matrices. The evaluation of $\nabla \Phi$ at the quadrature points involves contracting with the tensor $\boldsymbol{G}_\phi$, defined as:
\begin{equation*}
\boldsymbol{G}_\phi=\left[\begin{array}{l}
S_2 \otimes S_1 \otimes D_0 \\
S_2 \otimes D_1 \otimes S_0 \\
D_2 \otimes S_1 \otimes S_0
\end{array}\right]
\end{equation*}
Thus, the local operator $A^K$ becomes:
\begin{equation}
    A^K=\boldsymbol{G}_\phi^T W \boldsymbol{G}_\phi,
\end{equation}
where $W$ is the precomputed matrix of Jacobian and integration weights. 
Using sum-factorization techniques, the interpolation and integration operations for face integrals are of similar form~\cite{KronbichlerKormann19,witte2021fast}. Consider the evaluation of $\nabla \Phi$ at all quadrature points of a face in 3D with normal in $x1$ direction, we have
\begin{equation*}
\left[\begin{array}{c}
\partial_0 \Phi \\
\partial_1 \Phi \\
\partial_2 \Phi
\end{array}\right]=
\left[\begin{array}{l}
S_2 \otimes S_f \otimes D_0 \\
S_2 \otimes D_f \otimes S_0 \\
D_2 \otimes S_f \otimes S_0
\end{array}\right]\Phi
\end{equation*}
where the $1 \times N$ matrices $S_f$ and $D_f$ evaluate the shape functions and their first derivative on the respective boundary of the 1D reference cell.

\begin{figure}[tp]
\centering
\includegraphics[width=.65\textwidth]{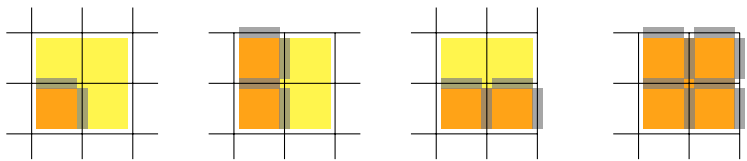}
\caption{Compute pattern for patch-wise integrals in 2D. Orange indicates cell integrals, while gary indicates face integrals. From left to right: center patch, top boundary patch, right boundary patch and corner patch.}
\label{fig:patch_int}
\end{figure}

In this study, focusing on tensor operations optimization, we consider a constant coefficient Cartesian grid. This allows us to integrate the cell and face integrals into a patch-based matrix-free method~\cite{Cui2024}. Then, the local operator $A_K$, can be rewritten as
\begin{equation}
\label{eq:tensorproduct_3d}
    A_K = L_2 \otimes M_1 \otimes M_0 + M_2 \otimes L_1 \otimes M_0 + M_2 \otimes M_1 \otimes L_0,
\end{equation}
where $L_d$ and $M_d$ are one dimensional stiffness matrices and mass matrices, respectively.
As demonstrated in Figure~\ref{fig:patch_int}, partial cell and face integrals in a patch are evaluated simultaneously. This patch-wise approach increases the dimensionality of local matrix operations, enabling efficient use of Tensor Cores even with low-order elements.

In the following benchmarks, we demonstrate the efficient evaluation of the finite element operator in the form of equation \eqref{eq:tensorproduct_3d}. It is important to note that the algorithm is specifically optimized for operations of the form $A \otimes B \otimes C$ acting on a given vector. This optimization makes our approach versatile, applicable to any operator constructed from a tensor product of one-dimensional entities.
This form of evaluation is analogous to performing a matrix-matrix multiplication, as shown in Appendix~\ref{sec:sf_code}, the dimensions involved are $N \times N$, which represent all the 1D basis functions evaluated at every 1D quadrature point or 1D mass (stiffness) matrix with special optimization considered on Cartesian mesh in~\eqref{eq:tensorproduct_3d}, and $N \times N^{d-1}$, accounting for the input values $u^{\left(i_1, \ldots, i_d\right)}$ and the intermediate results, see also~\cite{cantwell2011h,buis1996efficient}. 

\section{Programming Tensor Cores}

Tensor Cores, specialized computation units within NVIDIA GPUs, significantly accelerate matrix multiplication operations of the form $D = A \times B + C$, including in-place operations where $C = A \times B + C$. These cores are integral to high-performance computing, offering enhanced efficiency in processing complex matrix operations. 
The most straightforward method to utilize Tensor Cores is through NVIDIA's libraries like CUTLASS~\cite{cutlass} and cuBLAS~\cite{cublas}. These libraries facilitate the General Matrix Multiply (GEMM)  operation across different precisions, enabling large-scale matrix computations. For instance, CUTLASS provides an accessible interface for handling diverse matrix sizes and types, streamlining the application of Tensor Cores in complex computations.

The C++ Warp Matrix Functions and inline PTX instructions offer more granular control over Tensor Cores. The former is suited for standard operations, requiring less programming effort, while inline PTX instructions provide full access to Tensor Core features, beneficial for customized functions. The choice between the two hinges on the specific requirements of the task – C++ Warp Matrix Functions for simplicity and inline PTX instructions for flexibility and optimization.
Table~\ref{tab:A100_tensorcore} shows the supported input and output data types of Tensor Cores and their performance characteristics. Notably, the low precision FP16 data type, with 5 bits of exponent and 10 bits of mantissa, can be used as inputs to Tensor Cores in Ampere architectures. 
\begin{table}[tp]
    \caption{A100 Tensor Core Input / Output Formats and Performance vs. FP64 FFMA~\cite{A100}. The dimensions of the matrices are collectively described by the tuple $m \times n \times k$, where $A$ is an $m\times n$ matrix, $B$ is a $k\times n$ matrix, and $C$ and $D$ are $m\times n$ matrices.}
\begin{tabular}{lccc}
\toprule
Input Operands & FP64 & FP32 & FP16 \\
Accumulator & FP64 & FP32 & FP32 \\
TFLOPS/s & 19.5 & 19.5 & 312 \\
Speedup-factor vs. FFMA & 2$\times$ & 2$\times$ & 32$\times$ \\
$m \times n \times k$ & $8 \times 8 \times 4$ & $8 \times 8 \times 4$ & $16 \times 8 \times 16$ \\
\bottomrule
\end{tabular}
    \label{tab:A100_tensorcore}
\end{table}

Performing computations in shared memory can significantly improve performance by avoiding longer-latency, random global accesses~\cite{swirydowicz2019acceleration,KARAKUS2019380}. Therefore, our focus is on computations and optimization strategies using Tensor Cores via shared memory. Accordingly, all data is initially loaded from VRAM into shared memory before computation. While we do not emphasize VRAM access optimization in this paper, approaches such as pipeline usage or alternative global numbering formats might offer further enhancements.

\subsection{C++ Warp Matrix Functions}

When utilizing the C++ Warp Matrix Functions (referred to as \emph{WMMA API}) for matrix-matrix multiplication and addition ($C = A \times B + C$) on Tensor Cores, as illustrated in Listing~\ref{listing:wmma_api}, the process involves several key steps: an array of registers, known as \emph{Fragments}, are first initialized to store parts of matrices $A,B$ and $C$ in lines 2-4. Then, input matrices are copied from shared memory into fragments (as in lines 5-8). This step is crucial for preparing the data for processing by the Tensor Cores. Once the data is in fragments, the matrix multiplication and add are executed on the Tensor Cores (line 10). A warp, consisting of 32 threads, works collaboratively to perform the operation. This cooperative approach is essential for efficient use of the Tensor Cores. After the computation, the resulting $C$ fragment is stored back into shared memory (line 12). The WMMA API facilitates this process by providing the necessary functions and handling the fragment management.


\begin{figure}
\begin{lstlisting}[caption={A simple matrix-matrix multiplication in double precision on Tensor Cores using WMMA API.}, label={listing:wmma_api}]
__device__ void matmul(const double *A, const double *B, double *C) {
    fragment<matrix_a, 8, 8, 4, double, row_major> a_frag;
    fragment<matrix_b, 8, 8, 4, double, row_major> b_frag;
    fragment<accumulator, 8, 8, 4, double> c_frag;
    fill_fragment(c_frag, 0.0f);
    // load data from Shared Memory to the Register File
    load_matrix_sync(a_frag, A, ...);
    load_matrix_sync(b_frag, B, ...);
    // use Tensor Cores to compute the matrix computation
    mma_sync(c_frag, a_frag, b_frag, c_frag);
    // store result to Shared Memory
    store_matrix_sync(C, c_frag);
}
\end{lstlisting}
\end{figure}

The WMMA API streamlines basic Tensor Core operations, reducing programming complexity. The \texttt{wmma.load} function, in particular, manages the specialized input operand storage layout required by Tensor Cores. However, it is important to note that \texttt{wmma} instructions have their limitations. They can only access a subset of Tensor Core features and have stringent requirements regarding the data layout in shared memory. For instance, pointers such as \texttt{*A, *B, *C} must be 256-bit aligned and point to the first element of the matrix. Moreover, the elements within each matrix row or column need to be contiguous in memory. It is also critical that all threads in a warp call the function to ensure defined results.

\subsection{Inline PTX instructions}

\begin{figure}[tp]
\centering
\includegraphics[width=.75\textwidth]{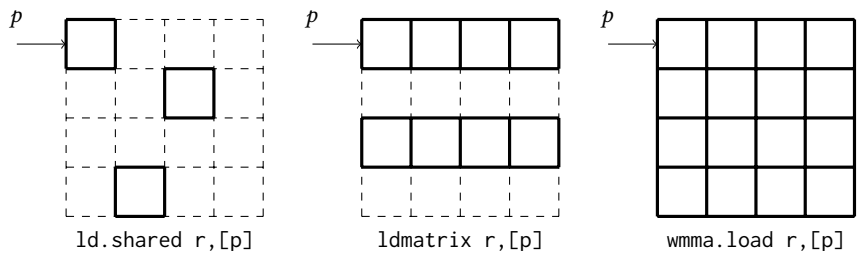}
\caption{Differences of three data movement instructions.}
\label{fig:data_ins}
\end{figure}

By contrast, the inline PTX instructions \texttt{mma.sync.aligned} (referred to as \emph{MMA instructions})~\cite{cudaPTX}, performing matrix multiply-and-accumulate computation, offer a more direct way to access all features of Tensor Cores compared to the WMMA API. This flexibility is crucial for achieving optimal performance, especially in complex operations that require fine-tuned control over data movement and processing.
Prior to executing Tensor Core instructions like \texttt{mma.sync.aligned}, it is necessary to manually load data from shared memory. Figure~\ref{fig:data_ins} illustrates three instructions for data movement. The generic per-thread \texttt{ld.shared} instruction takes a pointer ($p$) to an element in shared memory and loads it into a destination register ($r$). In a warp operation, 32 elements are loaded from shared memory, with each thread handling one element. On the other hand, \texttt{wmma.load} and \texttt{ldmatrix} operate on a warp-wide basis. All 32 threads in a warp work together to load the required data, enhancing efficiency and throughput.

Compared to WMMA API load instructions, which require the entire row (or column) elements of matrix stored consecutively, MMA instructions are more flexible. This flexibility, however, comes at the cost of additional index computations, potentially leading to some overhead. Nevertheless, one significant advantage of MMA instructions is their ability to utilize specialized data layouts to prevent bank conflicts in shared memory. This capability is particularly useful in scenarios where memory access patterns are non-standard or require optimization. In Listing~\ref{listing:mma_api}, we present the pseudocode of simple matrix-matrix multiplication in mixed precision using \texttt{mma} instructions. Unlike WMMA API shown in Listing~\ref{listing:wmma_api}, we explicitly distribute matrix elements across different threads in warp in lines 6-16. Then, the \texttt{mma.sync.aligned} instruction is employed to execute matrix multiplication on Tensor Cores. Finally, the results are explicitly stored back to shared memory.

\begin{figure}
\begin{lstlisting}[caption={A simple matrix-matrix multiplication in mixed precision on Tensor Cores using MMA instructions.}, label={listing:mma_api}]
__device__ void matmul(const half *A, const half *B, float *C) {
    float c[4] = {};
    half a[8];
    half b[4];
    // load data from Shared Memory to the Register File
    auto smem_Aptr = static_cast<uint32_t>(__cvta_generic_to_shared(&A));
    float *A_ptr = reinterpret_cast<float *>(&a);
    asm volatile("ldmatrix.sync.aligned.x4.m8n8.shared.b16 "
                 "{%0, %1, %2, %3}, [%4]; "
                 : "=f"(dA_ptr[0]), "=f"(dA_ptr[1]), "=f"(dA_ptr[2]),
                   "=f"(dA_ptr[3]) : "r"(smem_Aptr));
    auto smem_Bptr = static_cast<uint32_t>( __cvta_generic_to_shared(&B));
    float *B_ptr = reinterpret_cast<float *>(&b);
    asm volatile("ldmatrix.sync.aligned.x2.m8n8.shared.b16 "
                 "{%0, %1}, [%2]; " : 
                 "=f"(B_ptr[0]), "=f"(B_ptr[1]) : "r"(smem_Bptr));
    // use Tensor Cores to compute the matrix computation
    asm volatile("mma.sync.aligned.m16n8k16.row.col.f32.f16.f16.f32 "
                 "{%0,%1,%2,%3}, {%4,%5,%6,%7}, {%8,%9}, 
                 {%10,%11,%12,%13};\n"
                 : "=f"(c[0]), "=f"(c[1]), "=f"(c[2]), "=f"(c[3])
                 : "r"(A[0]), "r"(A[1]), "r"(A[2]), "r"(A[3]), 
                   "r"(B[0]), "r"(B[1]),
                   "f"(c[0]), "f"(c[1]), "f"(c[2]), "f"(c[3]));
    // store result to Shared Memory
    C[idx] = c[0];
    ...
}
\end{lstlisting}
\end{figure}

\subsection{Kernel design}

Recall that the tensor contraction operations for each mesh element can be executed independently. Thus, to parallelize the finite element operator we assign each element to a thread block on the GPU. Previous study~\cite{Cui2024} demonstrated that a finite element operator implementation using CUDA Cores could achieve close to 40\% peak performance. This forms our baseline for exploring acceleration using Tensor Cores. We adopt a 2D thread structure, allocating one thread per ``column" degree of freedom (DoF) for high-order elements in three dimensions. This design caters to the specific demands of high-order finite element calculations.
A critical aspect of our design involves the execution of instructions (both for data movement and computation) in warps. We propose a strategy where a single warp is responsible for processing multiple slices, or alternatively, multiple warps handle a single slice. This approach differs from traditional thread management and is tailored to leverage the unique capabilities of Tensor Cores.

Tensor Cores support specific matrix dimensions (as detailed in Table~\ref{tab:A100_tensorcore}). To demonstrate the efficiency of Tensor Cores with various data types, we present two benchmarks. These benchmarks are designed to highlight the effective use of Tensor Cores under different scenarios. In Section~\ref{sec:appendix}, we explore how matrices of arbitrary dimensions can be accelerated using Tensor Cores by applying the padding technique.

In our discussion, we refer to the matrix dimension $N$ instead of the polynomial degree $k$. For instance, $N = 8$ corresponds to $k = 7$ in cell-wise operations or $k = 3$ in patch-wise operations. This notation helps clarify the relation between matrix dimensions and polynomial orders in our Tensor Core operations.

\section{Double Precision benchmark}\label{sec:double_BK}

In this section, we focus on the acceleration of finite element operator evaluation in double precision. We explore the use of both the WMMA API and MMA instructions, supplemented by a series of targeted optimizations to enhance performance. In the following benchmarks, the whole Laplace operator, including face integrals and full data access, were benchmarked under the three-dimensional case, with problem sizes ranging from $10^3$ to $10^8$.

All numerical experiments in this work are performed on a single NVIDIA Ampere A100 SXM4 GPU, 1.27 GHz with 80GB of high-speed HBM2e memory for VRAM which provides up to 2TB/s peak memory bandwidth, hosted on a system with two AMD EPYC 7282 16-Core processors. Performance data for our experiments is gathered using NVIDIA's profiling tool \texttt{Nsight Compute}~\cite{cudaNcu}.

\subsection{Kernel optimizations}

Optimizing the local operator $A_K$, as defined in~\eqref{eq:tensorproduct_3d}, is crucial for efficient finite element computations. This core operation involves matrix multiplication of dimensions $N \times N$ and $N \times N^2$ 
in three dimensions. The input values for different directions necessitate ``reshape" operations. This is achieved by treating the 3D tensor as a matrix in row-, column-, or z-major format, with corresponding strides of $1, N, N^2$.

\textbf{\textit{WMMA Kernel}}. In this kernel, we use WMMA API to accelerate the matrix multiplication in direction $x0$ and $x1$. As the matrix elements within a row or column are required to be contiguous in memory, CUDA Core is used for computing direction $x2$. According to formula~\eqref{eq:tensorproduct_3d} it is evident that only 2/3 of the operations utilize Tensor Core.

\begin{figure}[tp]
\centering
\includegraphics[width=.75\textwidth]{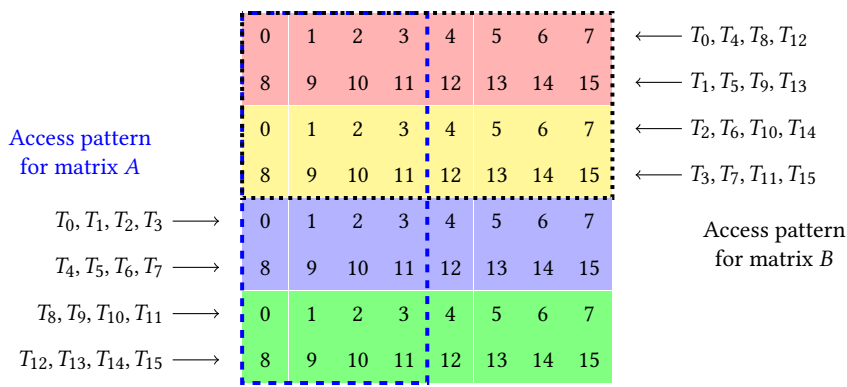}
\caption{Shared memory access pattern for matrix-matrix multiplications in double precision. The number in the grid indicates the bank number corresponding to the shared memory address. $T_i$ represents thread with index $i$.}
\label{fig:bank_comp}
\end{figure}

\textbf{\textit{MMA Kernel}}. In this kernel (referred to as \emph{MMA basic}), \texttt{ld.shared} instruction is used for data management, where there is no restriction on memory layout. After each thread having correct data in its own register file, \texttt{mma} instruction performs the matrix product with the entire warp. While all computations are accelerated, bank conflicts in shared memory access reduce the theoretical acceleration ratio, as depicted in Figure~\ref{fig:bank_comp}.

\begin{figure}[tp]
\centering
\includegraphics[width=.33\textwidth]{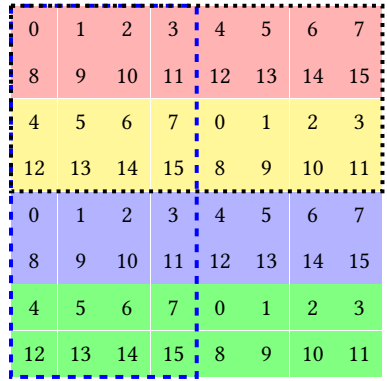}
\caption{Permuted shared memory data layout in double precision. Within each 4x4 sub-matrix, every thread accesses distinct bank numbers, adhering to a conflict-free access pattern.}
\label{fig:bank_comp_xor}
\end{figure}

\textbf{\textit{MMA Conflict Free Kernel}}. This kernel (referred to as \emph{MMA CF}) focuses on eliminating bank conflicts in shared memory. Figure~\ref{fig:bank_comp} illustrates the date layout with its corresponding bank number. For matrix $A$ at first phase, left half elements (blue dashed box) are loaded to the register, so this access pattern results in two-way bank conflicts. The same problem exists for matrix $B$, as shown in black dotted box. With the flexibility of loading instructions, A permuted data layout computed via bit-wise XOR operations, as shown in Appendix~\ref{sec:sf_code_tc}, is adopted to ensure conflict-free access patterns (Figure~\ref{fig:bank_comp_xor}). 

\begin{table}[tp]
    \centering
    \caption{Performance metrics for different memory types with CUDA Core implementation using single precision at $N=16$.}
    \begin{tabular}{lccc}
        \toprule
        & Shared & Constant & Texture* \\ 
        \midrule
        Duration (ms) & 3.83 & 4.31 & 8.62 \\
        \# of Regs. & 93 & 96 & 255 \\
        L1 \% Peak & 83.4 & 61.9 & 30.5 \\
        L1/TEX hit rate (\%) & 22.7 & 12.2 & 87.9 \\
        Pipe Lsu/Adu/Tex (\%) & 78.9/0.9/0 & 62.5/45.4/0 & 36.0/0.4/11.6 \\
        Warp stall & MIO Throttle (2.37) & MIO Throttle (2.39) & Tex Throttle (3.52) \\
        Blocks per SM & 2 & 2 & 1 \\
        \midrule
        \multicolumn{4}{l}{* 316 bytes spill stores, 564 bytes spill loads} \\
        \bottomrule
    \end{tabular}
    \label{tab:perf_shared_const_tex}
\end{table}

In addition to using shared memory, we explore alternative memory options for storing the one-dimensional matrices in~\eqref{eq:tensorproduct_3d}. Constant memory, though capable, encounters similar issues as bank conflicts in shared memory and strict address access requirements. Table~\ref{tab:perf_shared_const_tex} illustrates a rise in Adu pipe utilization from nearly 0\% to 45\%.
While the texture unit performs significantly better than random constant accesses, especially with address divergence in the warp, the kernel experiences register pressure that leads to register spilling, negatively impacting performance (Table~\ref{tab:perf_shared_const_tex}). Given that our data is small enough to fit in shared memory and its frequent access pattern, shared memory emerges as the optimal choice for achieving peak performance in this scenario.

Loop unrolling enhances performance by increasing scheduler flexibility and improving instruction-level parallelism, but it also introduces challenges, as noted in previous studies~\cite{Rawat2018}. In our context, where kernel performance heavily relies on register utilization, experiments reveal that unrolling inner loops often leads to register spills.
For instance, consider the \emph{MMA CF} kernel in the double-precision case with $N=16$. Initially using 216 registers per thread, achieving 12.5\% occupancy, with one active thread block per multiprocessor, unrolling the inner loop raises register usage to 254. This increase causes register spills and 16-byte overflow stores/loads\footnote{Register spilled can be indicated by flag ``-Xptxas -v''.}. Metrics from Nsight Compute profiler\footnote{We gathered the numbers with metrics: `smsp\_\_sass\_inst\_executed\_op\_local\_*.sum' for local load/store; `l1tex\_\_t\_sectors\_lookup\_*.sum' for L1 hit rate; `lts\_\_t\_sectors\_lookup\_*.sum' for L2 hit rate and `sm\_\_warps\_active.avg.pct\_of\_peak\_sustained\_active' for achieved occupancy.} show a rise in local store/load instructions to 476,656, with a 5\% increased in L1 and L2 hit rates. However, this increases memory traffic and instruction count, leading to a 10\% longer execution time.
In cases of half-precision computations without loop unrolling, each thread utilizes 128 registers, supporting 2 thread blocks per multiprocessor and achieving 25\% occupancy. Upon unrolling, register usage climbs to 147 per thread, reducing occupancy to one thread block per multiprocessor and extending execution time from 1.42ms to 2.02ms.

\subsection{Performance analysis}

\begin{figure}[tp]
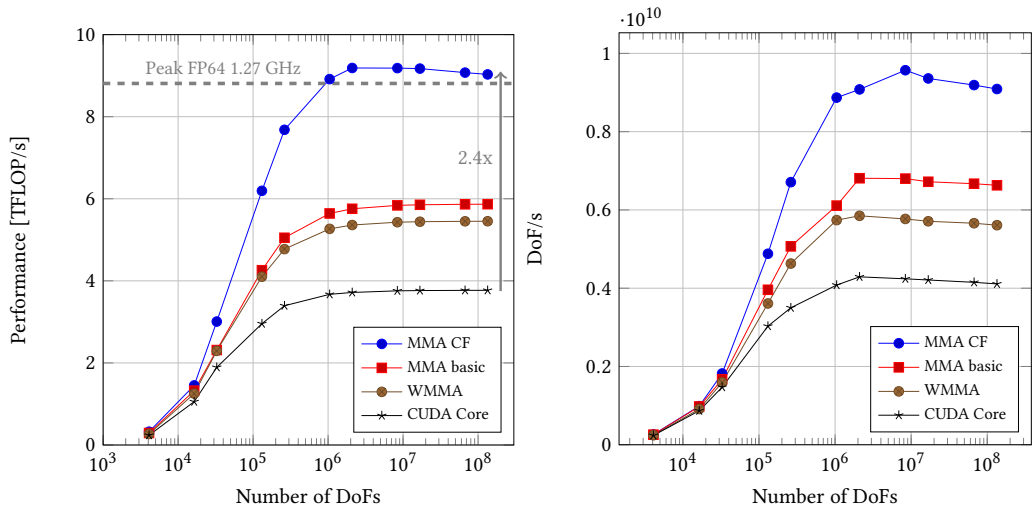

\centering
\includegraphics[width=.45\textwidth]{figure/double_perf.tikz}
\includegraphics[width=.45\textwidth]{figure/throughput_N8.tikz}
\caption{Arithmetic performance of implementation variants for finite element operator $Au$ with double precision in 3D for $N=8$.}
\label{fig:Arithmetic_d_8}
\end{figure}

The performance comparison of different kernels, measured in terms of Degrees of Freedom per second (DoF/s) and arithmetic throughput in TFLOPS/s, is illustrated in Figure~\ref{fig:Arithmetic_d_8}. A notable initial observation is that even a basic implementation using the WMMA API shows improvement over the original program, which was already highly optimized with CUDA Cores. This trend of enhancement becomes more pronounced when operations are fully accelerated using MMA instructions. The flexibility inherent in MMA instructions allows for more effective utilization of Tensor Core capabilities, resulting in further performance gains.
A key factor contributing to performance gains is the optimization of shared memory access patterns. The introduction of the Conflict-Free kernel significantly reduces wavefronts by nearly half, which in turn substantially decreases the stall times associated with MIO pipeline usage, from 11.58 to 2.11. The primary source of stalls in the optimized kernel is attributed to execution dependency, an expected outcome given the synchronization requirements in tensor operations across different directions. This aspect of the analysis confirms the high level of optimization achieved in the kernel. TFLOPS/s does not reveal the actual speed of operator evaluation. To gain a better understanding of algorithm performance, we also assess the throughput of the operations in DoF/s. The improvement in TFLOPS/s is also reflected in DoF/s as well, with similar behavior demonstrated, approaching 10 BDoF/s in the best case.

\begin{figure}[tp]
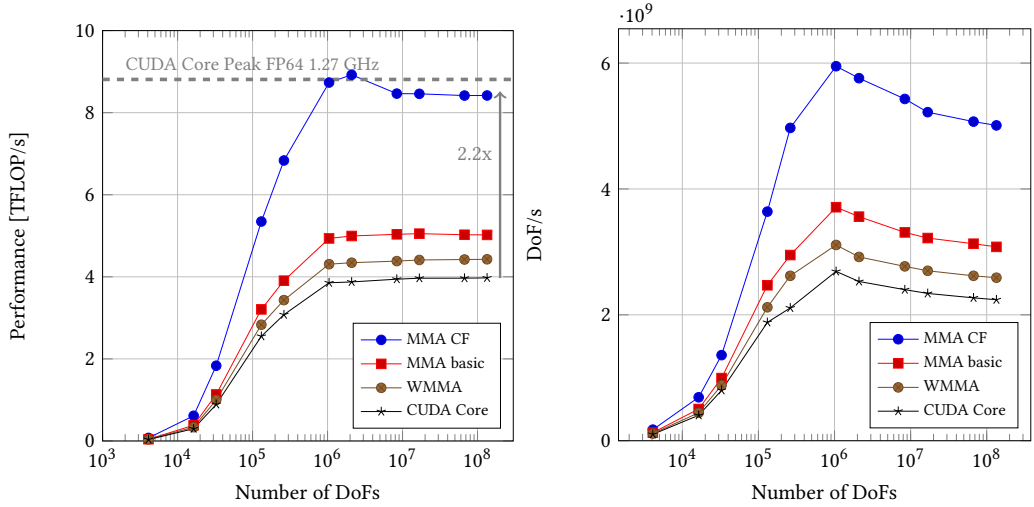

\centering
\includegraphics[width=.45\textwidth]{figure/double_perf_N16.tikz}
\includegraphics[width=.45\textwidth]{figure/throughput_N16.tikz}
\caption{Arithmetic performance of implementation variants for finite element operator $Au$ with double precision in 3D for $N=16$.}
\label{fig:Arithmetic_d_16}
\end{figure}

One remarkable finding from the analysis is that the most optimized kernel exceeds theoretical performance expectations, achieving a 2.3x improvement over the original CUDA Cores, which surpasses the anticipated 2x improvement. This additional speedup can be understood in terms of Arithmetic Intensity, or the concept of data reuse. In the context of matrix-matrix multiplication in device memory, submatrices from each input matrix are initially transferred to shared memory. Performing multiplication in shared memory capitalizes on data reusability, thereby reducing reliance on device memory. Similarly, when utilizing Tensor Cores, a comparable strategy is employed, albeit with matrices loaded into registers instead of shared memory. This method involves loading matrices into registers and then collectively computing within a warp. This heightened Arithmetic Intensity significantly contributes to the observed additional speedup.

\section{Half Precision benchmark}\label{sec:half_BK}

Drawing on the methodologies applied in the double-precision benchmark, similar optimization techniques have been implemented for half-precision computations. A significant difference in this context is the adjustment in data type size, which impacts bank numbers and necessitates a reevaluation of access patterns to effectively manage bank conflicts. Nevertheless, the fundamental strategy aligns with that used in the double-precision benchmark, as previously illustrated in Figure~\ref{fig:bank_comp_xor}). Additionally, to accommodate the half-precision format, wider loading instructions such as \texttt{ldmatrix} have been employed for reading data from shared memory.

\subsection{Performance analysis}

\begin{figure}[tp]
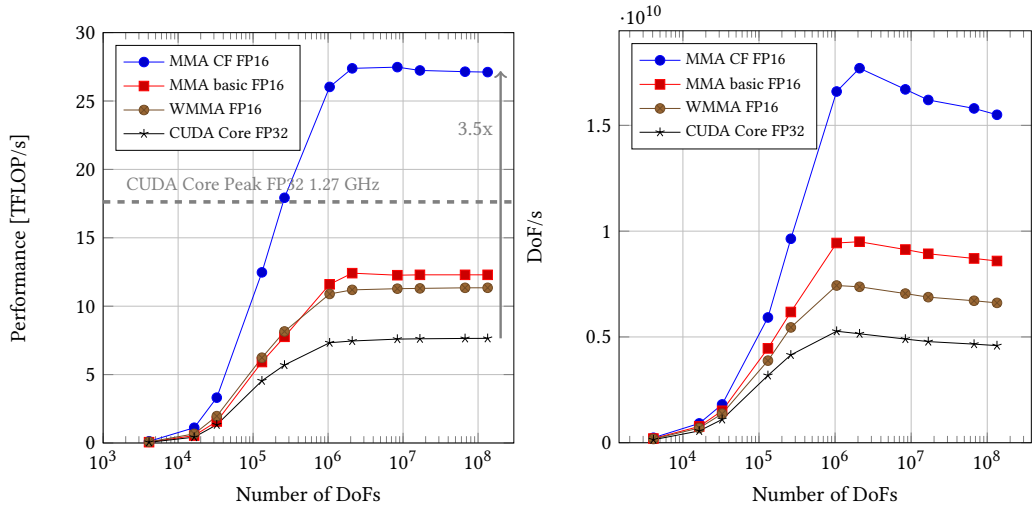

\centering
\includegraphics[width=.45\textwidth]{figure/single_perf.tikz}
\includegraphics[width=.45\textwidth]{figure/throughput_N16_s.tikz}
\caption{Arithmetic performance of implementation variants for finite element operator $Au$ with lower precision in 3D for $N=16$}
\label{fig:Arithmetic_s}
\end{figure}

The performance results of the half-precision operations utilizing Tensor Cores are presented in Figure~\ref{fig:Arithmetic_s}, showcasing similar behavior to the observations in the double-precision context (Figure~\ref{fig:Arithmetic_d_8}). 
Notably, the most efficient kernel in half-precision accelerates the original program using CUDA Cores by a factor of 3.5, surpassing even the theoretical performance of single-precision CUDA Cores. This improvement stems from reduced strain on the pipeline due to fewer and wider load instructions, coupled with the inherently higher peak performance capabilities of half precision.

\subsection{Precision loss}

Mixed-precision floating-point arithmetic and low-precision data types are gaining prominence as powerful tools for enhancing computational performance in Deep Learning applications and some HPC applications~\cite{gupta2015deep,goddeke2007performance}. While these approaches offer significant performance benefits, they invariably introduce a trade-off in terms of precision and accuracy, particularly when compared to traditional FP64/FP32 computations. 

The shift to lower precision, such as in Tensor Core operations, necessitates a thorough understanding of the resulting accuracy implications.
Previous studies~\cite{sun2022dissecting,markidis2018NVIDIA,ootomo2023reducing} mainly focus on analyzing the element-wise numerical behaviors and precision loss in matrix multiplications. Error correction methods have been explored to counteract the loss of precision, aiming to achieve the accuracy levels of FP32 calculations. We adopt the technique developed in~\cite{ootomo2023reducing} to mitigate rounding errors within Tensor Cores. The process of conducting single-precision matrix-matrix multiplication in half-precision, as represented in $C_{single} = A_{single}B_{single}$, is outlined in Algorithm~\ref{alg:matmul}.
\begin{algorithm}[tp]
\caption{Error Correction process for single-precision matrix-matrix multiplication $C_{single} = A_{single}B_{single}$ in half-precision.}\label{alg:matmul}
\begin{algorithmic}[1]
\State $A_{half} \leftarrow \texttt{float2half}(A_{single}) $ \Comment{convert precision}
\State $B_{half} \leftarrow \texttt{float2half}(B_{single}) $ 
\State $\delta A_{half} \leftarrow \texttt{float2half}((A_{single} - \texttt{half2float}(A_{half})) \times 2^{11}) $ \Comment{get residual}
\State $\delta B_{half} \leftarrow \texttt{float2half}((B_{single} - \texttt{half2float}(B_{half})) \times 2^{11}) $ 
\State $C_{single} \leftarrow A_{half}B_{half} + (\delta A_{half} B_{half} + A_{half}\delta B_{half}) / 2^{11} $ \Comment{output result}
\end{algorithmic}
\end{algorithm}

\begin{figure}[tp]
\centering
\includegraphics[width=.45\textwidth]{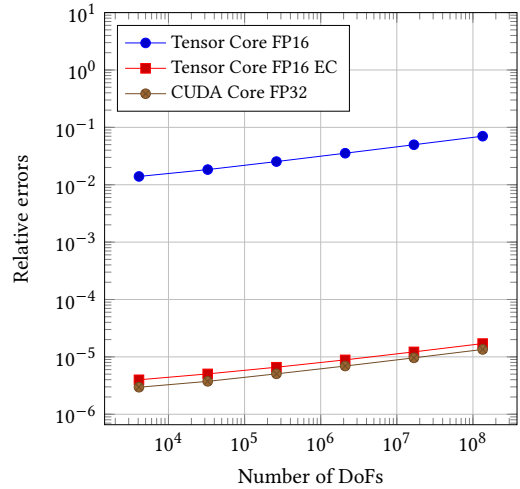}
\caption{Numeric profiling of matrix multiplication with different data types FP32 and FP16.}
\label{fig:error_prof}
\end{figure}

We use a simple application, the matrix-free evaluation of finite element operator $v = Au$, to evaluate the effect of low-precision FMA. The matrix-free evaluation is a typical computation pattern of iterative solvers which consists of many matrix-vector multiplications. The vector $u$ is randomly generated by normal distribution with $\mu = 0$ and $\sigma = 1$ in FP64, and we evaluate the relative $l_2$ errors between FP64 results $v^{FP64}$ and low-precision Tensor Core results $v^\ell$ using:
\begin{equation*}
    RelativeError = \frac{\sqrt{\sum_i(v^\ell_i - v^{FP64}_i)^2}}
    {\sqrt{\sum_i(v^{FP64}_i)^2}}.
\end{equation*}

\begin{figure}[tp]
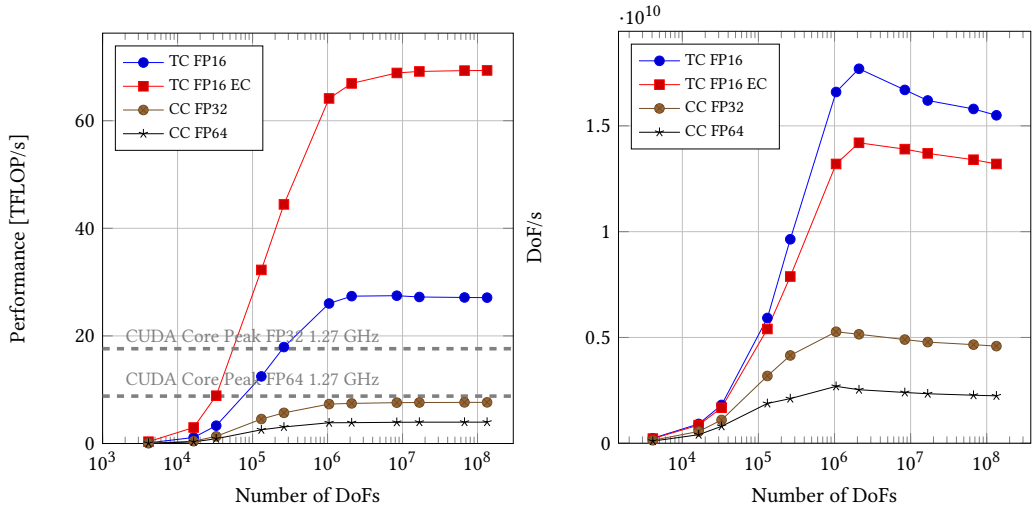

\centering
\includegraphics[width=.45\textwidth]{figure/error_p.tikz}
\includegraphics[width=.45\textwidth]{figure/error_perf.tikz}
\caption{Comparison of throughput for finite element operator $Au$ for $N=16$ with different data types using Tensor Cores(TC) and CUDA Cores(CC) in 3D.}
\label{fig:error_perf}
\end{figure}

Figure~\ref{fig:error_prof} shows the results of three types of matrix-vector multiplication with different problem size. First we notice that in general errors increase with the size of the vector. Then, the results of FP16 give much worse error level compared to using FP32 because of fewer mantissa bits. Using error correction strategy successfully recovers the same accuracy as obtained in FP32. Meanwhile, the additional overhead due to error correction is quantified in Figure~\ref{fig:error_perf}, measured in terms of throughput per DoF per second. Although data conversions between different precisions and additional matrix operations affect the efficiency of the algorithm, the utilization of Tensor Core still yields significantly notable results compared to traditional CUDA Cores.

The underlying rationale for this approach is rooted in the limitations of representing 32-bit values with 16-bit numbers, which inherently leads to precision loss. We address this by assigning the residual portion of the value, which is unrepresentable in 16 bits, to a separate 16-bit number. This strategy ensures that the original 32-bit value is fully represented by two 16-bit numbers, albeit with a minor error resulting from the assignment. This error correction method allows for compensating the precision loss incurred during input conversion by conducting supplementary operations on the residual value. Depending on the application's accuracy requirements, one can opt to refine one or both matrices, balancing the trade-offs among additional computational time, memory usage, and desired precision.

\subsection{Arbitrary matrix size}\label{sec:appendix}
As highlighted in Section~\ref{sec:mf_operator}, leveraging Tensor Cores for matrix multiplication imposes strict requirements on the dimensions of the matrices involved. When faced with matrices of non-standard dimensions, two primary approaches can be considered to still harness the acceleration capabilities of Tensor Cores.

The key to accommodating matrices of varying dimensions lies in the innovative use of the \texttt{mma} instruction, which necessitates the involvement of the entire warp in the computation. To address this, we can employ a padding strategy on the threads. This approach allows us to manipulate how data is loaded, utilizing the flexibility provided by the \texttt{ldmatrix} or \texttt{ld.shared} instructions.

Opting to have the padded threads not participate in the read operation presents a viable solution. This method has the advantage of completing the computation with additional threads while not requiring extra shared memory. However, this strategy is not without its challenges: One significant drawback is the potential for branch divergence. Our experimental findings indicate that this can have a substantial negative impact on computational efficiency. The irregular layout of memory banks corresponding to our data disrupts the regularity typically preferred in such computations. This irregularity makes it challenging to design an efficient new access pattern for each specific case, leading to potential bank conflicts.

\begin{figure}[tp]
\centering
\includegraphics[width=.45\textwidth]{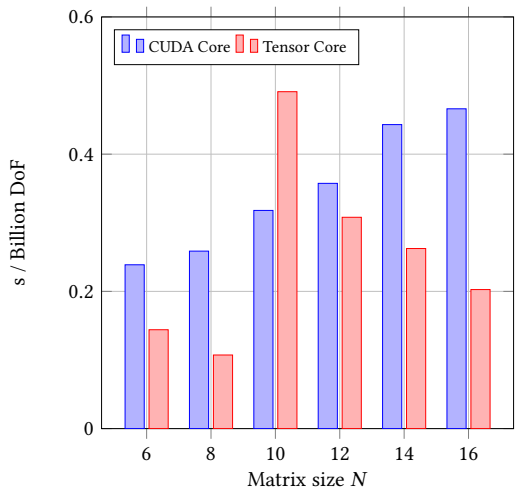}
\caption{Comparison of throughput for finite element operator $Au$ with arbitrary matrix size in 3D.}
\label{fig:all_perf}
\end{figure}

An effective alternative to managing varying matrix sizes involves padding both the memory and the threads. This strategy aligns with the optimizations previously discussed in Section~\ref{sec:double_BK}. Specifically, in three-dimensional computations, padding is required only in the $x0$ and $x1$ directions to ensure effective use of this strategy. Figure~\ref{fig:all_perf} compares computational throughput using CUDA Cores and Tensor Cores, with results for Tensor Cores reflecting the best-performing strategy discussed. This comparison reveals an expected trend: the more regular the matrix dimensions, the higher the efficiency of the computation. This effect is most pronounced within the matrix dimension range of 10 to 16. With threads and memory padded to comply with MMA instructions, matrix dimensions from 10 to 14 execute identical operations as $N = 16$. Consequently, the additional unnecessary operations lead to decreased performance, particularly noticeable at $N=10$. This result highlights a critical challenge in Tensor Core computations – the need for regularity in matrix dimensions to maintain high efficiency. When matrix dimensions become irregular or exceed specific sizes, the computational efficiency can suffer considerably.

\section{Roofline Analysis}

\begin{figure}[tp]
\centering
\includegraphics[width=.75\textwidth]{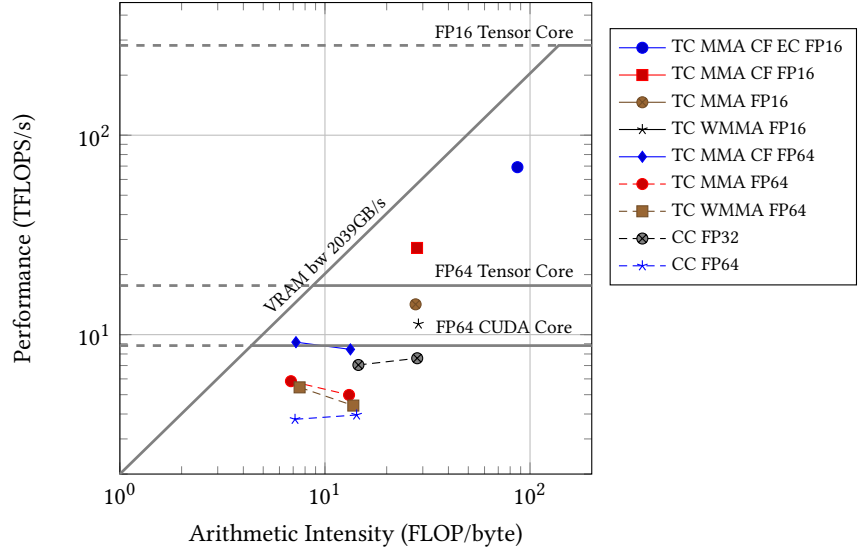}
\caption{VRAM roofline performance model for finite element operator $Au$ with various GPU implementations in 3D. Comparison of Tensor Cores (TC) 
and CUDA Cores (CC)
at different precisions for matrix size $N = 8, 16$ on NVIDIA A100 GPU at 1.27 GHz. Points connected by lines represent data $N = 8, 16$, from left to right. Discrete data points represent $N=16$ only, due to matrix size limitations of the MMA instructions.}
\label{fig:roofline}
\end{figure}

\begin{figure}[tp]
\centering
\includegraphics[width=.75\textwidth]{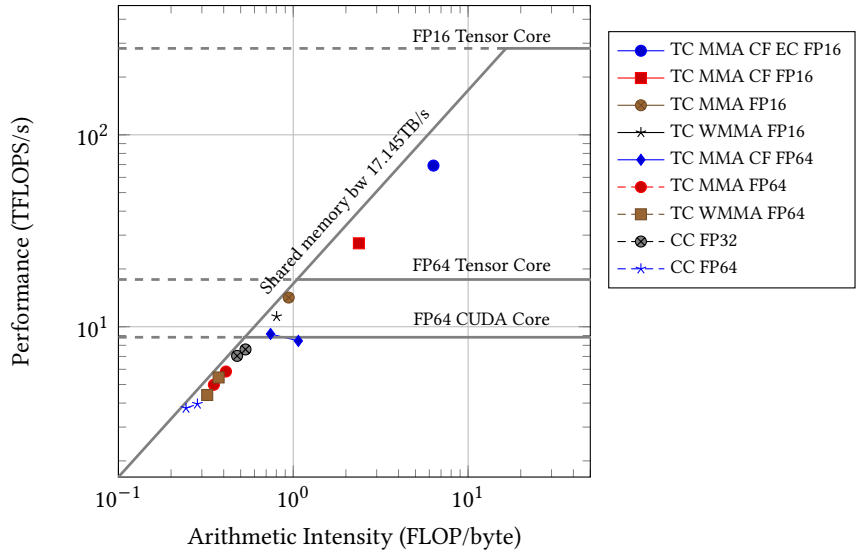}
\caption{Shared memory roofline performance model for finite element operator $Au$ with various GPU implementations in 3D. Comparison of Tensor Cores (TC) and CUDA Cores (CC) at different precisions for matrix size $N = 8, 16$ on NVIDIA A100 GPU at 1.27 GHz. Points connected by lines represent data $N = 8, 16$, from left to right. Discrete data points represent $N=16$ only, due to matrix size limitations of the MMA instructions.}
\label{fig:shmem_roofline}
\end{figure}

Figure~\ref{fig:roofline} displays the performance of various kernels in terms of the roofline performance model. Traditional CUDA Core implementations for matrix-free evaluation of the Laplacian operator are typically compute-bound.
However, the introduction of Tensor Cores has led to significant performance improvements, approaching the theoretical limits outlined by the roofline model. In the case of double precision, the optimal MMA CF kernel achieves 7.6 TFLOPS/s, which exceeds the peak performance of the corresponding CUDA Cores and lies between memory-bound and compute-bound. While for the single-precision case, all kernels are memory-bound. In particular, the TC MMA CF EC FP16 kernel employs error correction algorithms and additional operations to enhance arithmetic intensity, achieving a peak performance of 70 TFLOPS/s.

The conventional roofline model is under the assumption that the VRAM bandwidth is the limiting factor. However, due to the heavy use of shared memory, we believe that shared memory bandwidth is the real limiting factor.
Inspired by~\cite{swirydowicz2019acceleration}, we consider a roofline model that specifically focuses on the shared memory bandwidth, enabling a more accurate understanding of the factors affecting the performance.
The bandwidth of shared memory can be estimated using the formula:
\begin{equation*}
    B = \# \text{SMs} \times \# \text{banks} \times \text{word length} \times \text{clock speed}.
\end{equation*}
For NVIDIA A100 GPU, the corresponding bandwidth is $108\cdot 32 \cdot 4 \cdot 1.27=17.145 \text{TB/s}$, which aligns with the measured shared bandwidth of 127.7 bytes/clk/SM as reported in~\cite{sun2022dissecting}. Then, the shared memory roofline model is given by:
\begin{equation}
    \mathcal{R} = B \cdot \frac{F}{d_r+d_w},
\end{equation}
Where $d_r$ denotes the number of bytes read from shared memory, $d_w$ denotes the number of bytes written to shared memory, and $F$ denotes the number of floating point operations which generate the shared traffic of $d_r+d_w$ bytes. The position in the shared memory roofline plot in Figure~\ref{fig:shmem_roofline} reveals that the shared memory bandwidth is the limiting factor and also confirms the efficiency of the algorithm proposed in this work,  which approaches the upper bound. With one exception is the TC MMA CF FP64 kernel at $N=16$. This is due to the excessive dynamic shared memory usage, 143.36 Kbyte/block, leading only one thread block to be allocated on each multiprocessor, and the achieved occupancy decreases from 24.2\% at $N=8$ to 12.4\%. 
In our patch-wise approach integrating cells and faces , as shown in Figure~\ref{fig:patch_int}, meaningful evaluations occur for only a fraction of cell evaluations. Thus, the number of arithmetic operations per DoF for the 3D Laplacian at $N = 16$, shown in Table~\ref{tab:flop_byte}, is higher than the values reported in other studies~\cite{KronbichlerKormann19}. The number of operations introduced by the error correction (line 5 of Algorithm~\ref{alg:matmul}) can also be visualized as about three times higher.

\begin{table}[tp]
    \caption{Number of arithmetic operations per degree of freedom for evaluating the cell and face integrals of the 3D Laplacian for $N=16$ in patch-wise manner as described in Section~\ref{sec:mf_operator}.}
\begin{tabular}{lcccccc}
\toprule
& CC FP64 & CC FP32 & TC WMMA FP64 & TC MMA CF FP64 & TC MMA CF FP16 & TC MMA CF EC FP16 \\
\midrule
Flop/DoF & 1738 & 1843 & 1776 & 1702 & 1736 & 5361\\
\bottomrule
\end{tabular}
    \label{tab:flop_byte}
\end{table}

\section{Application: Geometric Multigrid method}

In this section, we embed the developed and optimized matrix multiplication kernel into a multigrid preconditioner with vertex-patch smoothers according to~\cite{Cui2024, witte2021fast}.

\subsection{The algorithm}

The V-cycle, see for instance~\cite{Bramble93,BrandtLivne11,Hackbusch85}, as the main component of the multigrid algorithm is described in Algorithm~\ref{alg:v_cycle}. The three basic operations of a V-cycle are 1) pre- and post-smoothing $S_\ell(x_\ell,b_\ell)$, 2) coarse grid solver$A_0^{-1}$, and 3) prolongation $I_{\ell-1}^\uparrow$ and restriction $I_{\ell-1}^\downarrow$. 
We use multigrid method with multiplicative vertex-patch smoothers~\cite{Cui2024, witte2021fast} as preconditioner for a flexible GMRES (FGMRES)~\cite{saad1993flexible} iterative solver. In the setting of mixed precision approach~\cite{goddeke2007performance,KronbichlerKormann19,KronbichlerLjunqkvist19,ruda2022}, the multigrid V-cycle is fully done in lower precision and the outer GMRES iteration is done in double precision. The format is converted when entering and exiting the V-cycle. Within the V-cycle, data is stored in single-precision format, which simplifies single-precision calculations by removing the need for data conversion. However, for half-precision calculations, on-the-fly precision conversion is necessary during computation. This is due to the error correction process and the accumulator format required by the \texttt{mma} instruction, which still relies on single-precision format. As shown in Table~\ref{tab:A100_tensorcore}, \texttt{mma} operations are conducted in mixed precision.

\begin{algorithm}[tp]
\caption{Multigrid V-cycle on level $\ell$.}\label{alg:v_cycle}
\begin{algorithmic}[1]
\Procedure{${x_\ell = \textsc{Vcycle}_\ell}$}{$A_\ell,x_\ell,b_\ell$}
\If{$\ell = 0$}
    \State \Return $x_0 \gets A_0^{-1}b_0$ \Comment{coarse grid solver}
\EndIf
\State $x_\ell \gets S_\ell(x_\ell,b_\ell)$ \Comment{pre-smoothing}
\State $r_\ell \gets b_\ell - A_\ell x_\ell$ \Comment{residual}
\State $r_{\ell-1} \gets I_{\ell-1}^\downarrow r_\ell$ \Comment{coarsen} 
\State $x_\ell \gets x_\ell
    + I_{\ell-1}^\uparrow \textsc{Vcycle}_{\ell-1}(A_{\ell-1},0,r_{\ell-1})$ \Comment{coarse grid correction}
\State \Return $x_\ell \gets S_\ell(x_\ell,b_\ell)$ \Comment{post-smoothing}
\EndProcedure
\end{algorithmic}
\end{algorithm}

\subsection{Poisson problem}

In this experiment, we delve into the practical application of low-precision computations, specifically focusing on solving the Poisson equation model problem as defined in equation \eqref{eq:poisson}. The right-hand side $f$ and the Dirichlet boundary data $g$ in~\eqref{eq:poisson} are configured to yield an analytical solution of
\begin{equation*}
    u(x,y,z)=\sin(\pi x)\sin(\pi y)\sin(\pi z).
\end{equation*}
The FGMRES iteration is stopped once the $l_2$ norm of the residual has decreased by $10^{-8}$ compared to the initial $l_2$ norm of the residual.

\begin{figure}[tp]
\centering
\includegraphics[width=.55\textwidth]{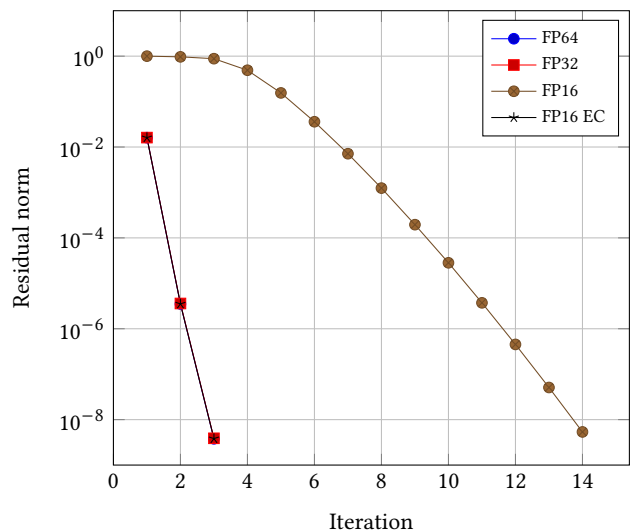}
\caption{Residual norm as a function of iteration steps for solving the Poisson equation with polynomial degree $k=7$ ($N = 16$) in 3D. Multigrid preconditioner running in different precision for FGMRES solver with relative accuracy of $10^{-8}$. The lines representing FP64, FP32, and FP16 EC completely overlap with each other.}
\label{fig:fgmres_residual}
\end{figure}

In Figure~\ref{fig:fgmres_residual}, we compare the number of iterations required for the FMGRES method using multigrid preconditioner at different precision. By employing an error correction technique, we successfully recover the same accuracy as obtained in double or single precision. Table~\ref{tab:fgmres} provides detailed solving times and error norms, highlighting the impact of precision on the algorithm's efficiency. While single precision offers high performance, its accuracy limitations necessitate a larger number of iterations for convergence, thus affecting the overall efficiency. Conversely, the algorithm augmented with error correction maintains effectiveness in accelerating the solution process, despite the additional matrix operations it requires.

\begin{table}
    \caption{FGMRES experiments with multigrid preconditioner in different precision for Poisson problem in 3D.}
    \begin{tabular}{lcclccclc}
    \toprule
         & \multicolumn{4}{c}{$\mathbb{Q}_3$, 16 MDoFs} & \multicolumn{4}{c}{$\mathbb{Q}_3$, 134 MDoFs} \\
    \cmidrule(lr){2-5} \cmidrule(lr){6-9} 
         & time [s] & \# its.  & $L_2$ error & Speedup & time [s] & \# its.  & $L_2$ error & Speedup \\
    \midrule
       FP64 CC  & 0.405 & 3 & $1.62 \times 10 ^{-9}$ & --- & 3.229 & 3 & $1.78 \times 10 ^{-10}$ & --- \\
       FP32 CC  & 0.227 & 3 & $4.86 \times 10 ^{-8}$ & 1.79 & 1.798 & 3 & $4.83 \times 10 ^{-8}$ & 1.79 \\
       \textbf{FP64 TC} &  0.215 & 3 & $1.62 \times 10 ^{-9}$ & 1.88 & 1.691 & 3 & $1.78 \times 10 ^{-10}$ & 1.91 \\
    \midrule
         & \multicolumn{4}{c}{$\mathbb{Q}_7$, 16 MDoFs} & \multicolumn{4}{c}{$\mathbb{Q}_7$, 134 MDoFs} \\
    \cmidrule(lr){2-5} \cmidrule(lr){6-9} 
         & time [s] & \# its.  & $L_2$ error & Speedup & time [s] & \# its.  & $L_2$ error & Speedup \\
    \midrule
       FP64 CC & 0.717 & 3 & $2.94 \times 10 ^{-12}$ & --- & 5.901 & 3 & $3.43 \times 10 ^{-12}$ & --- \\
       FP32 CC &  0.374 & 3 & $1.46 \times 10 ^{-12}$ & 1.92 & 3.044 & 3 & $2.18 \times 10 ^{-12}$ & 1.94 \\
       \textbf{FP64 TC} & 0.369 & 3 & $2.94 \times 10 ^{-12}$ & 1.94 & 3.025 & 3 & $3.43 \times 10 ^{-12}$ & 1.95 \\
       FP16 TC &  0.588 & 14 & $4.31 \times 10 ^{-11}$ & 1.22 & 9.733 & 29 & $4.32 \times 10 ^{-11}$ & 0.60 \\
       \textbf{FP16 TC EC} &  0.167 & 3 & $1.04 \times 10 ^{-11}$ & 4.29 & 1.322 & 3 & $3.32 \times 10 ^{-11}$ & 4.46 \\
    \midrule
    \multicolumn{9}{l}{* CC: CUDA Core; TC: Tensor Core; EC: Error Correction; \# its.: Number of iteration steps.} \\ 
    \bottomrule
    \end{tabular}
    \label{tab:fgmres}
\end{table}

\begin{figure}[tp]
\centering
\includegraphics[width=.55\textwidth]{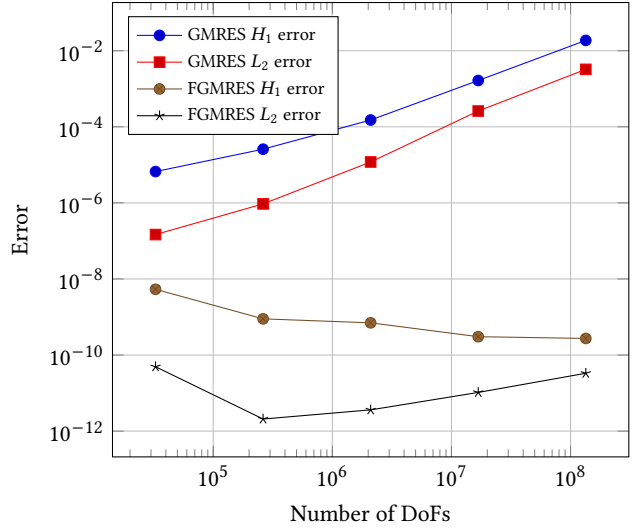}
\caption{Residual norm as a function of iteration steps for solving the Poisson equation with polynomial degree $k=7$ ($N = 16$) in 3D. Multigrid preconditioner running in different precision for FGMRES solver with relative accuracy of $10^{-8}$. The lines representing FP64, FP32, and FP16 EC completely overlap with each other.}
\label{fig:fgmres_vs_gmres}
\end{figure}

Our choice of the relatively complex FGMRES method is based on two main considerations. First, the simple Conjugate Gradient (CG) method or flexible CG method is unsuitable for the multiplicative Schwarz smoother used in the multigrid method. This is because parallelized execution of the smoother results in a non-symmetric preconditioner. Second, the truncation error introduced by using half-precision computations necessitates the robustness of FGMRES, which allows for different preconditioners in each iteration step. This capability makes FGMRES more resilient to inaccuracies in preconditioner evaluations.
As shown in Figure~\ref{fig:fgmres_vs_gmres}, the relative error of the FGMRES method remains  stable, at $10^{-10}$. In contrast, the $L_2$ error and $H_1$ error of the GMRES method increase with mesh refinement, underscoring the benefits of FGMRES in maintaining accuracy.

\section{Conclusion}
This paper has presented a comprehensive approach to accelerating the solution of the Poisson problem using Tensor Cores, a technology with increasing relevance in large-scale real-world applications. We have detailed strategies to optimize GPU kernels for finite element operators, focusing on reducing memory pipeline pressure and resolving bank conflicts.

Our investigation underscores the superiority of MMA instructions over the WMMA API when programming Tensor Cores. The benchmarks outlined in Sections \ref{sec:double_BK} and \ref{sec:half_BK} have demonstrated the distinct benefits of using the \texttt{mma} instructions. A critical advantage of MMA instructions is their ability to eliminate bank conflicts entirely, thanks to the flexibility offered by the load instruction.
The most finely tuned kernel in the double precision benchmark reaches an impressive 8 TFLOPS/s, approximately 45\% of the peak FP64 performance of Tensor Cores. This performance represents a 2.3-fold increase in speed over a highly optimized CUDA Core implementation. Notably, this level of efficiency is maintained across matrix sizes $N=8$ and $N=16$. 
In half precision, we achieved a 3.5-fold speedup, surpassing the theoretical peak performance of FP32. This result underscores the potential of Tensor Cores to significantly enhance computational efficiency in existing applications.

We perform a detailed comparison on a real application --- the Poisson problem, that highlights the advantages of Tensor Core over traditional CUDA Core in mixed-precision computation. On the one hand, accelerating double-precision operations with Tensor Core results in half the solution time, which is even more pronounced at larger problem sizes.
Moreover, employing low-precision multigrid methods for preconditioning leads to a more than fourfold increase in solution efficiency while upholding accuracy and convergence rates. Thus, while maximizing the utilization of Tensor Cores necessitates inline PTX instructions and adherence to specific matrix dimensions, prioritizing the development of algorithms compatible with Tensor Core is crucial, as it facilitates substantial computational acceleration.

Beyond the advancements presented in this work, further research needs to be carried out on developing efficient algorithm on other differential operators and finite elements, such as Raviart–Thomas elements used in Stokes problems. 
Efficient utilization of Tensor Cores on non-standard matrix multiplications, especially with anisotropic tensor product elements, remains a critical area of exploration. 
Additionally, vector-valued problems introduce more degrees of freedom, necessitating optimization in the process of loading data from global memory to shared memory, which is another focus for future work.

\section*{Acknowledgement}

The author thanks Guido Kanschat, Heidelberg University, for valuable discussions and comments on the manuscript. Also, the author would like to thank China Scholarship Council (CSC NO. 202106380059) for the financial support.

\bibliographystyle{ACM-Reference-Format}
\bibliography{references}

\appendix
\section{CUDA Core implementation for evaluation of Laplacian operator by sum factorization in 3D}\label{sec:sf_code}

In Listing~\ref{listing:sum_factorized_cc}, we provide partial code examples illustrating the multiplication of the local vector $u_K$ with, e.g. the matrix $M_2\otimes M_1\otimes L_0$, as presented in~\eqref{eq:tensorproduct_3d}, for three-dimensional cases. To avoid the naive $\mathcal{O}(N^{2d})$ arithmetic cost, the Kronecker product matrix is not applied in its expanded $N^d \times N^d$ form. Instead, each of the $d$ factors is processed separately in a rearranged manner known as sum factorization. Initially, the multiplication starts with $L_0$ (\texttt{shape\_data}) and $u_K$ (\texttt{in}). This operation involves multiplying the $N\times N$ (\texttt{n\_dofs\_1d}) matrix $L_0$ with the $N\times N^2$ matrix obtained by reshaping $u_K$ into column-major form. As shown in line 9-12, reshaping is accomplished by altering the indices. Similarly, for directions 1 and 2 correspond to different ways of indexing. Matrices $M_i$ or $L_i$, always keep the row-major forms (line 7). In lines 6 and 15, a for loop is used to traverse the $N$ matrix multiplications because of the two-dimensional thread block structure with size $N\times N$.


\begin{figure}
\begin{lstlisting}[caption={Device code for evaluation of Laplacian operator in 3D using CUDA Cores.}, label={listing:sum_factorized_cc}]
template <typename Number, int n_dofs_1d, int direction>
__device__ void apply(const Number *shape_data, const Number *in, Number *out) {
    ...
    Number pval[n_dofs_1d] = {};
    const unsigned int stride = n_dofs_1d * n_dofs_1d;
    for (unsigned int z = 0; z < n_dofs_1d; ++z) 
        for (unsigned int k = 0; k < n_dofs_1d; ++k) {
            const unsigned int shape_idx = row * n_dofs_1d + k;
            const unsigned int source_idx =
                (direction == 0) ? (col * n_dofs_1d + k + z * stride) :
                (direction == 1) ? (k * n_dofs_1d + col + z * stride) :
                                   (z * n_dofs_1d + col + k * stride);
            pval[z] += shape_data[shape_idx] * in[source_idx];
        }
    for (unsigned int z = 0; z < n_dofs_1d; ++z) {
        const unsigned int destination_idx =
            (direction == 0) ? (col * n_dofs_1d + row + z * stride) :
            (direction == 1) ? (row * n_dofs_1d + col + z * stride) :
                               (z * n_dofs_1d + col + row * stride);
        out[destination_idx] = pval[z];
    }
    ...
}
\end{lstlisting}
\end{figure}

\section{Tensor Core implementation for evaluation of Laplacian operator by sum factorization in 3D}\label{sec:sf_code_tc}

In Listing~\ref{listing:sum_factorized_tc}, we present the Tensor Core implementation corresponding to the CUDA Core implementation shown in Listing~\ref{listing:sum_factorized_cc}, focusing on achieving an optimal conflict-free access pattern. Since different precision and matrix sizes correspond to different \texttt{mma} instructions, here we take $N=8$ in double precision as an example. The reshaping operation when using Tensor Core is similarly achieved by adopting different indexing methods, as shown in line 8, where the matrix is considered in row-major format. The key step to avoid bank conflict is the permutation of the indexes by the XOR operation (line 9) to achieve the conflict free access pattern. Similarly this approach is applied to matrix $u_k$ (lines 13-15). Subsequently the matrix multiplication accumulation operation on the Tensor Core is executed using the \texttt{mma} instruction (line18-22).
For the sake of simplicity, we show the implementation for direction 1 here, the approach for other directions follows a similar pattern.

\begin{figure}
\begin{lstlisting}[caption={Device code for evaluation of Laplacian operator in 3D using Tensor Cores with $N=8$ in double precision.}, label={listing:sum_factorized_tc}]
template <typename Number, int n_dofs_1d, int direction>
__device__ void apply(const Number *shape_data, const Number *in, Number *out) {
    ...
    if (direction == 0) {...}
    else if (direction == 1) {
        double2 c[n_dofs_1d / 2] = {};
        for (int cycle = 0; cycle < 2; ++cycle){
            const int a_idx = (row * n_dofs_1d + col + cycle * 4) ^
                            Util::get_base<n_dofs_1d>(row, 0);
            auto a0 = shape_data[a_idx];

            for (int z = 0; z < n_dofs_1d / 2; ++z){
                const int b_idx =
                    ((col + cycle * 4) * n_dofs_1d + row + (z * 2 + warpId) * offset) ^
                    Util::get_base<n_dofs_1d>(col + cycle * 4, z * 2 + warpId);
                 auto b0 = in[b_idx];

                asm volatile(
                    "mma.sync.aligned.m8n8k4.row.col.f64.f64.f64.f64 "
                    "{%0,%1}, {%2}, {%3}, {%4,%5};\n"
                    : "=d"(c[z].x), "=d"(c[z].y)
                    : "d"(a0), "d"(b0), "d"(c[z].x), "d"(c[z].y));
                }
            }
        for (int z = 0; z < n_dofs_1d / 2; ++z){
            const int c_idx =
                (row * n_dofs_1d + 2 * col + (z * 2 + warpId) * offset) ^
                Util::get_base<n_dofs_1d>(row, z * 2 + warpId);
              *((double2 *)(out + c_idx)) = c[z];
            }
    }
    else if (direction == 2) {...}
    ...
}
\end{lstlisting}
\end{figure}

\end{document}